\documentclass[reprint,a4paper,aps,pra,amsmath,amsfonts,floatfix,nofootinbib,longbibliography]{revtex4-2}
\usepackage{graphicx}
\usepackage{xcolor}
\usepackage{bm}
\usepackage{dcolumn}
\usepackage{fixmath}

\usepackage[bookmarks=false]{hyperref}
\hypersetup{colorlinks=true,linktoc=all,linkcolor=blue,citecolor=blue,	citebordercolor=blue}
\hypersetup{pdfstartview=FitH}
\hypersetup{pdfstartpage=1}

\newcommand{\vps}[1]{\varphi_{#1}}
\newcommand{\be}{\vec{e}}
\newcommand{\bA}{\bm{A}}
\newcommand{\braket}[3]{\left\langle#1\left|\rule{0pt}{1pt} #2\right|#3\right\rangle}
\renewcommand{\ni}{\vec{\nabla}_{\!1}}
\newcommand{\nj}{\vec{\nabla}_{\!2}}
\newcommand{\ev}[1]{\left\langle#1\right\rangle}
\newcommand{\oh}{\frac{1}{2}}
\newcommand{\br}{{\vec{r}}}
\newcommand{\bR}{{\vec{R}}}
\newcommand{\riA}{r_{1A}}
\newcommand{\rjA}{r_{2A}}
\newcommand{\riB}{r_{1B}}
\newcommand{\rjB}{r_{2B}}
\newcommand{\rij}{r_{12}}
\newcommand{\rAB}{r_{AB}}
\newcommand{\me}{m_\mathrm{e}}
\newcommand{\mun}{\mu_\mathrm{n}}
\newcommand{\icm}{\mathrm{cm}^{-1}}
\newcommand*{\centt}[1]{\multicolumn{1}{c}{#1}}
\newcommand*{\cent}[1]{\multicolumn{1}{c}{$#1$}}
\newcolumntype{x}[1]{D{.}{.}{#1}}
\newcommand{\nn}{\nonumber\\}
\newcommand{\etal}{{\em et al.}}
\newcommand{\ASen}{\mathcal{E}^{(5)}_\mathrm{AS,en}}

\graphicspath{{Figs/}}

\begin{document}

\title{
Electron-nucleus Araki-Sucher correction in the hydrogen molecule isotopologues}

\author{Jacek Komasa} 
\email{komasa@man.poznan.pl}
\affiliation{Faculty of Chemistry, Adam Mickiewicz University,
             Uniwersytetu Poznańskiego 8, 61-614 Pozna{\'n}, Poland}

\date{\today}

\begin{abstract}
The quantum electrodynamic Araki-Sucher correction arising from the interaction between electrons and nuclei is calculated for rovibrational energy levels of the hydrogen molecule and its isotopologues. The corresponding expectation value $\ev{r_{en}^{-3}}$ is evaluated across a wide range of internuclear distances using the Born-Oppenheimer approximation. The electronic wave function is represented as a linear combination of explicitly correlated Gaussian basis functions. This correction contributes approximately tenths of a megahertz (about $10^{-5}\, \icm$) to the dissociation energy of rovibrational levels and to the transitions between them. Given recent spectroscopic measurements with an accuracy of 10 kHz, this correction is necessary to achieve sub-MHz agreement between theory and experiment.
\end{abstract}

\maketitle

\section{Introduction}

High-precision spectroscopic measurements of rovibrational transitions in molecular hydrogen and its isotopologues \cite{Fast:20,Diouf:20,Fast:22,Cozijn:23,Diouf:24,Cozijn:24,Stankiewicz:25} have reached a level of accuracy that necessitates the inclusion of subtle quantum electrodynamical effects in theoretical models. Due to their simplicity and the availability of accurate wave functions, these systems serve as ideal platforms for testing the limits of molecular quantum electrodynamics (QED).

Among the radiative corrections contributing to the energy levels of light molecules, the Araki–Sucher term, originating from the long-range part of the Breit interaction, plays a significant role. This term accounts for the exchange of transverse photons and exhibits a logarithmic divergence in the interparticle coalescence limit. Its proper treatment requires regularization techniques and careful numerical evaluation. The Araki–Sucher term, initially formulated in the context of atomic QED \cite{Araki:57,Sucher:58}, has since been extended to molecular systems \cite{PCK:05,Korobov:06}, where its contribution becomes non-negligible in high-precision theoretical predictions. Previous theoretical treatments of molecules have primarily focused on the dominant interelectron Araki–Sucher correction, with numerical values estimated mainly within the Born-Oppenheimer approximation 
\cite{Piszczatowski:09,PK:10,Komasa:11,Balcerzak:17,Stanke:17,SPKP:23}.
An exception is the direct nonadiabatic calculation for the ground energy level of all isotopologues
of the hydrogen molecule \cite{PKCP:19,PKSP:19}.

In this work, we present a detailed computation of the electron-nucleus Araki–Sucher correction for the rovibrational states of the hydrogen molecule and all its isotopologues. The calculations are performed within the Born–Oppenheimer approximation, employing the perturbative QED method and accurate electronic wave functions. Our results quantify the contribution of this term to transition energies and assess its relevance to current and anticipated spectroscopic experiments. This study contributes to the ongoing effort to refine theoretical predictions for simple molecular systems and to support the interpretation of precision measurements at the sub-MHz level \cite{PK:25a}.

\section{Theoretical fundamentals}

Our calculations are based on the nonrelativistic quantum electrodynamic (NRQED) theory, where  the energy of a bound rovibrational level of a light molecule is expanded in powers of the fine-structure constant $\alpha$ 
\begin{align}
E(\alpha)&=\alpha^2m\,E^{(2)}+\alpha^4m\,E^{(4)}
 +\alpha^5m\,E^{(5)}\nn
&\quad+\alpha^6m\,E^{(6)}+\alpha^7m\,E^{(7)}+\dots\,. \label{Eq:alpha}
\end{align}
Explicit expressions for the expansion coefficients $E^{(i)}$, which are valid for systems like H$_2$, are well documented (see, for example,~Ref.~\onlinecite{PKCP:19}) and will not be included here. In this work, we concentrate on a specific component of the QED correction ($\alpha^5m\,E^{(5)}$), known as the Araki-Sucher correction.

\subsection{Definitions}

For H$_2$, the Araki-Sucher correction takes the following form (in a.u.)
\begin{align}\label{Eq:E5ASdef}
E^{(5)}_\mathrm{AS}&=
-\frac{7}{6\,\pi}\sum_{a\neq b}\frac{\me^2}{m_a\,m_b}\ev{\frac{1}{r_{ab}^3}},
\end{align}
where $a$ and $b$ index all four particles. From now on, $\me$ will denote the electron mass, while the nuclear masses will be represented as $m_A$ and $m_B$. 
The expectation value above denotes the following limit:
\begin{align}
	\bigg\langle \frac{1}{r_{ab}^{3}} \bigg\rangle &\equiv 
\lim_{\varepsilon\rightarrow0}\left[\left\langle\frac{\Theta(r_{ij}-\varepsilon)}{r_{ab}^3}\right\rangle +
(\gamma_E+\ln \varepsilon)\big\langle 4 \pi \delta( \vec r_{ab})\big\rangle\right],\label{EQ:ASdef}
\end{align}
where $\gamma_E$ is the Euler-Mascheroni constant, and $\Theta(x)$ 
is the Heaviside step function.

The formula~(\ref{Eq:E5ASdef}) can be split according to different powers of the electron-to-nucleus fraction
\begin{align}
E^{(5)}_\mathrm{AS}&=
 -\frac{7}{6\,\pi}\!\left(\!\ev{\frac{1}{r_{12}^3}}+\frac{2\,\me}{\mun}\!\ev{\frac{1}{r_{1A}^3}}
 + \frac{\me^2}{m_A\,m_B}\!\ev{\frac{1}{r_{AB}^3}}\!\right)\!,
\end{align}
where $\mun=m_A\,m_B/(m_A+m_B)$ is the reduced nuclear mass. 
The leading term, the electron-electron Araki-Sucher correction, has recently been evaluated using the BO approximation by Siłkowski \etal~\cite{SPKP:23}. Of interest to us is the second term, the electron-nucleus Araki-Sucher correction. Due to the mass factor, this term is 2-3 orders of magnitude smaller than the first one. The contribution from the nucleus-nucleus correction is another three orders of magnitude smaller and is therefore neglected for the time being.

\subsection{Regularization}

Gaussian-type wave functions, which we use in this work, inherently fail to satisfy the interparticle cusp conditions. This limitation results in a slow convergence of expectation values for nearly singular operators, such as $\ev{r_{1A}^{-3}}$. To address this issue, we implement a regularization method that significantly improves the convergence of the relevant expectation values, thereby enhancing the accuracy of the Araki-Sucher correction.
In this context, the expectation value $\ev{r_{1A}^{-3}}$ can be represented as follows \cite{PCK:05,SPKP:23}
\begin{align}\label{Eq:reg}
\ev{\frac{1}{r^3_{1A}}}
&= 2\,\mu_{1A}\!\left[
 \left(1\!+\!\gamma_E \right)\!\Bigl(2\,{V}_{1A}\!-\!{R}_{1A}\Bigr) 
 \!+\!\Bigl( 2\, \tilde{V}_{1A}\!-\!\tilde{R}_{1A} \Bigr)
\right]\!.
\end{align}
In this equation, $\mu_{1A}=\me\,m_A/(\me+m_A)$ is the reduced atomic mass. Furthermore,
\begin{align}
{V}_{1A} &\equiv \ev{\frac{1}{r_{1A}}\,(E - V)}, \\
\tilde{V}_{1A} &\equiv \ev{\frac{\ln r_{1A}}{r_{1A}}\,(E - V)}, \\
{R}_{1A}&\equiv-\sum_{i=1,2}\ev{\vec\nabla_i\frac{1}{r_{1A}}\,\vec \nabla_i}, \\
\tilde{R}_{1A}&\equiv-\sum_{i=1,2}\ev{\vec\nabla_i\frac{\ln r_{1A}}{r_{1A}}\,\vec \nabla_i}\nn
&\quad -\sum_{X=A,B}\frac{1}{m_X}\ev{\vec\nabla_X\frac{\ln r_{1A}}{r_{1A}}\,\vec \nabla_X},
\end{align}
where 
\begin{align}
V&=\frac{1}{\rij}+\frac{Z_A\,Z_B}{\rAB}-\frac{Z_A}{\riA}-\frac{Z_B}{\riB}-\frac{Z_A}{\rjA}-\frac{Z_B}{\rjB}
\end{align}
is the Coulomb potential of the system and $E$ is an eigenvalue of the molecular Hamiltonian.
In our calculations, the last term in $\tilde{R}_{1A}$ was neglected due to the additional small mass factor. To maintain consistency with the BO approximation, we also assume that $\mu_{1A}=1$.

\subsection{Wave function}

The total wave function is approximated as a product of electronic and nuclear wave functions
\begin{align}\label{Eq:adwf}
\Psi(\br_1,\br_2,\bR_A,\bR_B)&=\Phi(\br_1,\br_2)\,\chi(\bR)\,,
\end{align}
where $\br_1$ and $\br_2$ represent the positions of the electrons, while $\bR_A$ and $\bR_B$ denote the positions of the nuclei, with the origin located halfway between the two nuclei. We also introduce $\bR=\bR_A-\bR_B$, a vector joining the nuclei. 

All the operators considered here are free of the nuclear derivatives. Therefore, the separation of electronic and nuclear variables, as introduced in Eq.~(\ref{Eq:adwf}), allows us to express the expectation value of an operator $Q$ in a nested form
\begin{align}
\ev{Q}\equiv\braket{\Psi}{Q}{\Psi}=\braket{\chi}{\braket{\Phi}{Q}{\Phi}}{\chi}.
\end{align}
In this expression, the inner bracket denotes integration over the electronic variables and depends parametrically on $\bR$. Conversely, the outer bracket represents integration over nuclear variables.

The electronic function $\Phi$ is a solution to the Schrödinger equation $H\Phi=\mathcal{E}_{\rm el}\Phi$ with the clamped nuclei Hamiltonian
\begin{align}\label{Eq:Ham}
H&=\oh\left(p_1^2+p_2^2\right) + V\,.
\end{align} 
$\Phi$ is represented as a linear combination of explicitly correlated Gaussian (ECG) functions:
\begin{align}
\Phi(\br_1,\br_2)&=\frac{1}{4}\left(1+\hat{P}_{12}\right)\left(1+\hat{P}_{AB}\right)\sum_k c_k\,\vps{k}(\br_1,\br_2)\,.
\end{align}
The operators $\hat{P}_{12}$ and $\hat{P}_{AB}$ exchange electrons and nuclei, respectively, ensuring proper symmetry of the wave function. The ECG basis functions are defined as~\cite{Singer:60}:
\begin{align}
\vps{k}(\br) &=\exp[-(\br-\vec{s}_k)\cdot\bA_k\cdot(\br-\vec{s}_k)]\nn
&=\exp[-\,\br\cdot\bA_k\cdot\br + 2\,\be_k\cdot\br - G_k]\,.
\end{align}
In this formula, the following symbols are introduced: $\br=(\br_1,\br_2)$, $\vec{s}_k=(\vec{s}_{k,1},\vec{s}_{k,2})$, $\be_k= \bA_k\cdot\vec{s}_k$, and $G_k=\vec{s}_k\cdot\bA_k\cdot\vec{s}_k$. Each basis function has its own set of variational parameters collected in the vector $\vec{s}_k$ and the matrix $\bA_k$. 

The calculations were performed for over 80 internuclear distances.
The ECG basis set was optimized for each internuclear separation to ensure that the energy converges to a precision better than $10^{-9}$ a.u. This set of electronic wave functions $\Phi$ and corresponding energies $\mathcal{E}_{\rm el}$ has been used previously to evaluate adiabatic, nonadiabatic, relativistic, and QED corrections~\cite{Komasa:11}. In the present study, these wave functions were applied to evaluate the potential of the electron-nucleus Araki-Sucher correction 
\begin{align}\label{Eq:ASenpot}
\ASen(R)&=-\frac{7}{3\,\pi}\frac{\me}{\mun}\braket{\Phi}{\frac{1}{r_{1A}^3}}{\Phi}
 +\mathcal{O}\left[\left(\frac{\me}{\mun}\right)^2\right].
\end{align}
The method for numerically calculating the $\ASen(R)$ potential is the main topic of this work and will be detailed in the following section.

The rovibrational energy levels $E_{v,J}$ and wave functions $\chi_{v,J}$ were determined by solving the radial Schrödinger equation with the $\mathcal{E}_{\rm el}$ potential
\begin{eqnarray}
\biggl[-\frac{1}{2\,R^2}\,\frac{\partial}{\partial R}\,
\frac{R^2}{\mu_{\rm n}}\,\frac{\partial}{\partial R}\,
+\frac{J\,(J+1)}{2\,\mu_{\rm n}\,R^2}&& 
\nonumber \\ 
+\mathcal{E}_\mathrm{el}(R)-E_{v,J}\biggr]\,\chi_{v,J}(R) &=& 0\,. \label{Eq:rse}
\end{eqnarray}
Finally, the electron-nucleus Araki-Sucher correction to the energy of a rovibrational level $(v,J)$ was calculated as an expectation value of the $\ASen(R)$ potential in the state $\chi_{v,J}$
\begin{align}\label{Eq:ASenaver}
E^{(5)}_\mathrm{AS,en}(v,J)&=\braket{\chi_{v,J}(R)}{\ASen(R)}{\chi_{v,J}(R)}.
\end{align}

\section{Matrix elements for $\bm\ASen$}

To compute the $\left\langle r_{1A}^{-3} \right\rangle$ of Eq.~(\ref{Eq:ASenpot}), we need to evaluate the four primary expectation values introduced in Eq.~(\ref{Eq:reg}): $V_{1A}$, $R_{1A}$, $\tilde{V}_{1A}$, and $\tilde{R}_{1A}$, with the wave function $\Phi$. When we explicitly insert the Coulomb potential operator, these expectation values can be expressed in terms of a set of 16 elementary expectation values
\begin{align}
{V}_{1A}
&=\left(\mathcal{E}_\mathrm{el}-\frac{1}{\rAB}\right)\ev{\frac{1}{r_{1A}}}
 -\ev{\frac{1}{r_{1A}}\frac{1}{\rij}}+\ev{\frac{1}{\riA^2}}\nn
&\quad 
 +\ev{\frac{1}{r_{1A}}\frac{1}{\riB}}
 +\ev{\frac{1}{r_{1A}}\frac{1}{\rjA}}
 +\ev{\frac{1}{r_{1A}}\frac{1}{\rjB}},\\
{R}_{1A}
&=-\ev{\ni\frac{1}{r_{1A}}\ni}-\ev{\nj\frac{1}{r_{1A}}\nj},\\
\tilde{V}_{1A}
&=\left(\mathcal{E}_\mathrm{el}-\frac{1}{\rAB}\right)\ev{\frac{\ln r_{1A}}{r_{1A}}}
 -\ev{\frac{\ln r_{1A}}{r_{1A}}\frac{1}{\rij}}\nn
&\quad +\ev{\frac{\ln r_{1A}}{r_{1A}^2}}
 +\ev{\frac{\ln r_{1A}}{r_{1A}}\frac{1}{\riB}}
 +\ev{\frac{\ln r_{1A}}{r_{1A}}\frac{1}{\rjA}}\nn
&\quad+\ev{\frac{\ln r_{1A}}{r_{1A}}\frac{1}{\rjB}},\\
\tilde{R}_{1A}&
 =-\ev{\ni\frac{\ln r_{1A}}{r_{1A}}\,\ni}
 -\ev{\nj\frac{\ln r_{1A}}{r_{1A}}\,\nj}.
\end{align}
In this section, we present a technique for evaluating these elementary expectation values. 

We recall two types of Gaussian integral transforms (for $n>0$)
\begin{align}
	\frac{1}{r^n} &= \frac{2}{\Gamma(n/2)} \int_0^{\infty} dt\,t^{n-1}\,e^{-r^2 t^2},\\
\frac{\ln r}{r^n} &= -\frac{1}{\Gamma(n/2)}\,\int_0^{\infty} dt\,t^{n-1}\left(2\ln t -\psi(n/2)\right)\,e^{-r^2 t^2},\ 
\end{align}
where $\psi(z)=\Gamma'(z)/\Gamma(z)$ is the digamma function.
By applying a pertinent transform to the elementary expectation value, we obtain an integral over $t$ that contains a primitive expectation value of the Gaussian-type operator. These primitive expectation values can be readily evaluated in the ECG basis. The infinite integration domain can be mapped to the finite $(0,1)$ interval using two different substitutions
\begin{align}\label{Eq:subs}
t&=\frac{1-x}{x}\,,\ dt=-\frac{1}{x^2}\,dx\,,& x&=\frac{1}{t+1}\,,\ (0,\infty)\to(1,0)\,, \\
t&=\frac{y}{1-y}\,,\ dt=\frac{1}{(1-y)^2}\,, & y&=\frac{t}{t+1}\,,\ (0,\infty)\to(0,1)\,.
\end{align}
Typically, one of the alternative mappings is better suited to meet the requirements of the numerical integration method used in the GAUSEXT subroutine \cite{GAUSEXT}. This subroutine is dedicated to the numerical integration of functions that have a logarithmic singularity at one end of the interval. It is essential to consider this characteristic when selecting a particular mapping.

Below we present examples of the application of the Gaussian integral transforms:
\begin{widetext}
\begin{align}\label{Eq:exevbeg}
\ev{\frac{\ln r_{1A}}{r_{1A}^2}}&=-\int_0^{\infty}\!\!dt\,t\,(2\ln t+\gamma_E)\ev{e^{-r_{1A}^2 t^2}}, \\
\ev{\frac{\ln r_{1A}}{r_{1A}}\frac{1}{r_{12}}}&=-\frac{1}{\sqrt{\pi}}\,\int_0^{\infty}\!\!dt\,(2\ln t + \gamma_E + \ln 4)\ev{e^{-r_{1A}^2 t^2}\frac{1}{r_{12}}},\\
\ev{\ni\,\frac{\ln r_{1A}}{r_{1A}}\,\ni}
&=-\frac{1}{\sqrt{\pi}}\int_0^{\infty}\!\!dt\,(2\ln t + \gamma_E + \ln 4)
 \ev{\ni\,e^{-r_{1A}^2 t^2}\,\ni}.\label{Eq:exevend}
\end{align}
\end{widetext}
The above primitive expectation values are among the most challenging and slowest-converging quantities within the entire 16-element set. Therefore, they were selected to assess the efficiency of numerical integration using the GAUSEXT program.

\section{Numerical calculations}

To estimate the performance of the integration method described above, we examined the convergence of the primitive expectation values from Eqs.~(\ref{Eq:exevbeg})-(\ref{Eq:exevend}) with increasing integration grid density. The results shown in Table~\ref{Tab:evconv} demonstrate that 5 to 6 significant digits can be reliably obtained.
\begin{table}[ht]
\caption{Convergence of exemplary expectation values of Eqs.~(\ref{Eq:exevbeg})-(\ref{Eq:exevend}) (in a.u.) with increasing integration grid size, $n$ (H$_2$, $R=1.4$ a.u., $K=1200$).}
\label{Tab:evconv}
\begin{center}
\begin{tabular*}{0.48\textwidth}{c@{\extracolsep{\fill}}*{3}{x{2.8}}}
\hline
$n$ & \cent{\ev{\dfrac{\ln r_{1A}}{r_{1A}^2}}} & \cent{\ev{\dfrac{\ln r_{1A}}{r_{1A}}\dfrac{1}{r_{12}}}} & \cent{\ev{\ni\dfrac{\ln r_{1A}}{r_{1A}}\ni}} \\
\hline
16 & -1.822\,989\,7 & -0.151\,781\,9 & 0.100\,146\,7 \\
32 & -1.823\,013\,0 & -0.151\,768\,5 & 0.100\,126\,3 \\
64 & -1.823\,014\,6 & -0.151\,767\,6 & 0.100\,124\,9 \\
\hline
\end{tabular*}
\end{center}
\end{table}

Another source of numerical uncertainty arises from the finite size of the ECG basis set, $K$. Table~\ref{Tab:ASconv} allows us to assess the error in $\ASen$ caused by both the grid size and the basis set size. The numerical errors associated with these two factors are comparable and are two orders of magnitude smaller than the uncertainty introduced by the neglected finite nuclear mass effects $\mathcal{O}\left[\left(\me/\mun\right)^2\right]$. For instance, at $R=1.4$ a.u., the latter is estimated as $\me/\mun\,\ASen=0.000\,001\,6$. Hence, the final BO value is
\begin{align}
\ASen(1.4)=0.001\,479\,8(16)\,.
\end{align}

\begin{table}[ht]
\caption{Convergence of $\ASen$ (in a.u.) with respect to both increasing basis set size, $K$, and increasing integration grid size, $n$ (H$_2$, $R=1.4$ a.u.).}
\label{Tab:ASconv}
\begin{center}
\begin{tabular*}{0.48\textwidth}{c@{\extracolsep{\fill}}*{3}{x{2.10}}}
\hline
$K\backslash n$ & \cent{16} & \cent{32} & \cent{64} \\
\hline
 300 & 0.001\,479\,574 & 0.001\,479\,698 & 0.001\,479\,712 \\
 600 & 0.001\,479\,643 & 0.001\,479\,766 & 0.001\,479\,778 \\
1200 & 0.001\,479\,653 & 0.001\,479\,774 & 0.001\,479\,784 \\
\hline
\end{tabular*}
\end{center}
\end{table}

\begin{figure}[hb]
\includegraphics[scale=0.4]{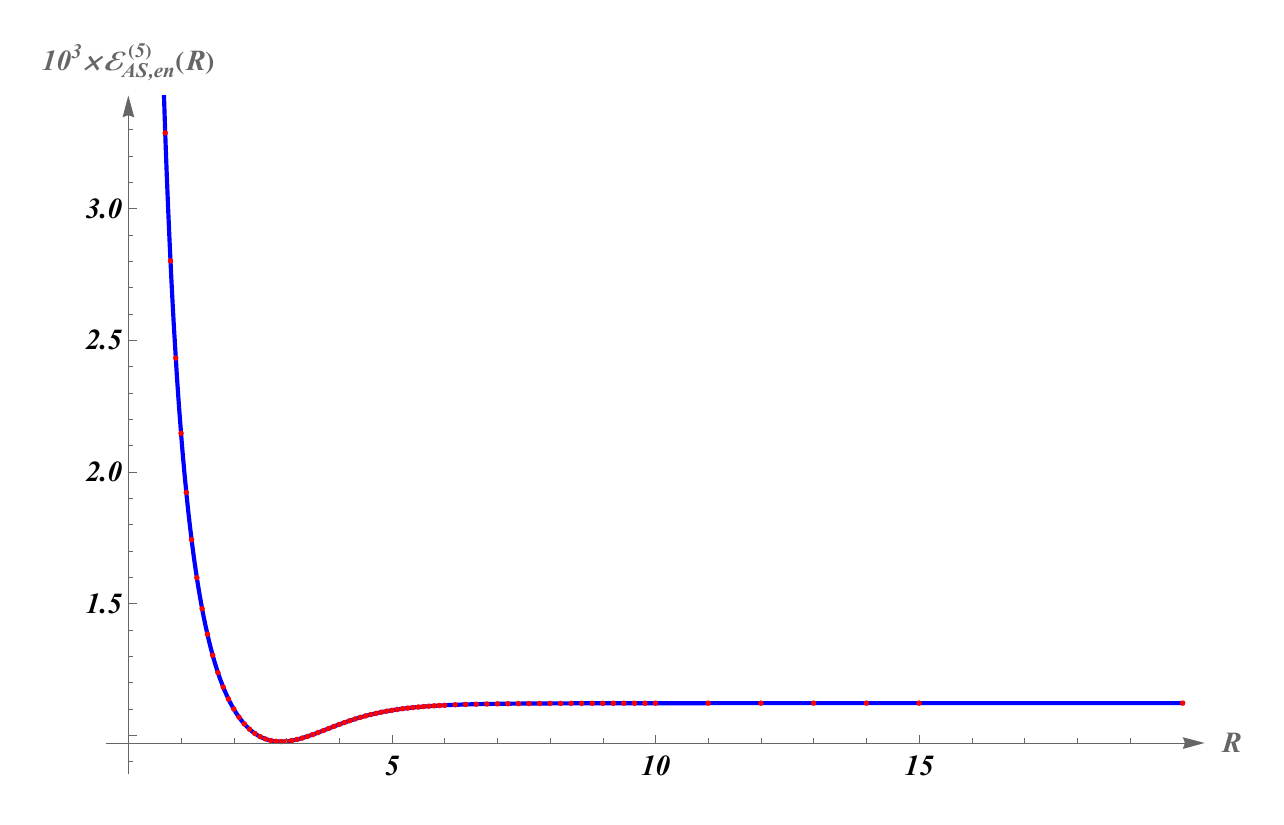}
\caption{Plot of the $\ASen(R)$ potential (in a.u.). The curve exhibits minimum at $R=2.894$ a.u.
and a horizontal asymptote at $\frac{14\,\ln{2}}{6\,\pi\,\mun}\approx 0.001121604$ a.u.}
\label{Fig:ASenpot}
\end{figure}

The values of $\ASen(R)$, calculated with $K=1200$ and $n=64$ across a wide range of internuclear distances, are presented in Table~\ref{Tab:ASenpot} and illustrated graphically in Figure~\ref{Fig:ASenpot}. This function features a single minimum at $R=2.894$ a.u. As $R$ approaches 0 from this minimum, $\ASen(R)$ increases sharply to a finite value of united atom. Conversely, as $R\to\infty$, the potential rises monotonically toward a horizontal asymptote determined by the limit of separated atoms
\begin{align}
\ASen(\infty)
& =\frac{14}{3\pi}\frac{\me}{\mu_n}\ln{2} =0.001121604\,\mathrm{a.u.}
\end{align}

\begin{table}[htb]
\caption{The electron-nucleus Araki-Sucher potential (in a.u.) calculated for H$_2$. For any other isotopolog $AB$, this potential must be rescaled by the appropriate ratio of the reduced nuclear masses $\mun(\mathrm{H}_2)/\mun(AB)$.}
\label{Tab:ASenpot}
\begin{tabular*}{0.48\textwidth}{*{3}{rx{2.9}@{\hspace{2ex}}}}
\hline\hline 
  $R\ $ &  \cent{\ASen(R)} &  $R\ $ &  \cent{\ASen(R)} &  $R\ $ &  \cent{\ASen(R)}   \\
\hline
  0.00 &  $-$         &  2.90 &  0.000975316 &   5.80 &  0.001111025 \\
  0.10 &  0.015626806 &  3.00 &  0.000976386 &   5.90 &  0.001112187 \\
  0.20 &  0.010926417 &  3.10 &  0.000979158 &   6.00 &  0.001113211 \\
  0.30 &  0.008022848 &  3.20 &  0.000983361 &   6.20 &  0.001114902 \\
  0.40 &  0.006129966 &  3.30 &  0.000988733 &   6.40 &  0.001116214 \\
  0.50 &  0.004844578 &  3.40 &  0.000995026 &   6.60 &  0.001117229 \\
  0.60 &  0.003941043 &  3.50 &  0.001002004 &   6.80 &  0.001118017 \\
  0.70 &  0.003286783 &  3.60 &  0.001009450 &   7.00 &  0.001118633 \\
  0.80 &  0.002800754 &  3.70 &  0.001017168 &   7.20 &  0.001119112 \\
  0.90 &  0.002431659 &  3.80 &  0.001024980 &   7.40 &  0.001119489 \\
  1.00 &  0.002146009 &  3.90 &  0.001032738 &   7.60 &  0.001119786 \\
  1.10 &  0.001921322 &  4.00 &  0.001040313 &   7.80 &  0.001120030 \\
  1.20 &  0.001742131 &  4.10 &  0.001047607 &   8.00 &  0.001120221 \\
  1.30 &  0.001597555 &  4.20 &  0.001054546 &   8.20 &  0.001120375 \\
  1.40 &  0.001479784 &  4.30 &  0.001061070 &   8.40 &  0.001120510 \\
  1.50 &  0.001383110 &  4.40 &  0.001067151 &   8.60 &  0.001120613 \\
  1.60 &  0.001303296 &  4.50 &  0.001072769 &   8.80 &  0.001120703 \\
  1.70 &  0.001237145 &  4.60 &  0.001077922 &   9.00 &  0.001120776 \\
  1.80 &  0.001182219 &  4.70 &  0.001082618 &   9.20 &  0.001120842 \\
  1.90 &  0.001136628 &  4.80 &  0.001086874 &   9.40 &  0.001120900 \\
  2.00 &  0.001098894 &  4.90 &  0.001090710 &   9.60 &  0.001120947 \\
  2.10 &  0.001067844 &  5.00 &  0.001094154 &   9.80 &  0.001120989 \\
  2.20 &  0.001042537 &  5.10 &  0.001097236 &  10.00 &  0.001121024 \\
  2.30 &  0.001022209 &  5.20 &  0.001099985 &  11.00 &  0.001121169 \\
  2.40 &  0.001006229 &  5.30 &  0.001102425 &  12.00 &  0.001121256 \\
  2.50 &  0.000994068 &  5.40 &  0.001104595 &  13.00 &  0.001121311 \\
  2.60 &  0.000985273 &  5.50 &  0.001106514 &  14.00 &  0.001121363 \\
  2.70 &  0.000979447 &  5.60 &  0.001108207 &  15.00 &  0.001121384 \\
  2.80 &  0.000976236 &  5.70 &  0.001109705 &  20.00 &  0.001121438 \\
\hline\hline 
\end{tabular*}
\end{table}

The data presented in Table~\ref{Tab:ASenpot} were interpolated using cubic splines. The long-range behavior was modeled using the expression
\begin{align}
\ASen(R)&=-\frac{7}{3\,\pi}\frac{\me}{\mun}\left(-2\ln{2}+\frac{1}{2\,R^3}+\frac{108.547}{R^6}\right),
\end{align}
which was obtained by fitting to the data points with $R\geq 9.0$ a.u. The interpolated and long-range curves were smoothly joined at $R=10.28$ a.u. This potential will be incorporated into the nearest release of the publicly available H2Spectre program.~\cite{H2Spectre}.

Rovibrational averaging of the $\ASen$ potential (see Eq.~(\ref{Eq:ASenaver})) was performed numerically using a highly dense integration grid. This process introduces no significant numerical error. The calculations were conducted for all six isotopologues of the hydrogen molecule across several of the lowest vibrational and rotational quantum numbers, specifically $v=0,\dots,3$ and $J=0,\dots,6$. The nuclear masses (in a.u.), recommended by CODATA 2022~\cite{CODATA:22}, were used to evaluate the nuclear reduced mass $\mun$.
The numerical values of the correction to dissociation energy are compiled in Table~\ref{Tab:ASendiss}. 

The largest absolute correction to the dissociation energy occurs for the ground state of all isotopologues, and this value decreases with increasing quantum numbers $(v,J)$ and with heavier nuclear masses. For the ground level of H$_2$, this correction is $-28.677\cdot10^{-6}\,\icm$ ($\approx-0.86$ MHz), corresponding to a relative contribution of $8 \cdot 10^{-10}$ to the total dissociation energy of $36\,118.069\,630(26)\,\text{cm}^{-1}$~\cite{PK:25a}. Although this correction is minor in absolute terms, its inclusion is essential for achieving the final theoretical uncertainty of $26 \cdot 10^{-6}\,\text{cm}^{-1}$ ($\approx 0.78$ MHz). 

The significance of the electron-nucleus Araki–Sucher term is also evident in precise determinations of transition frequencies. For instance, one of the most precisely measured molecular transition frequencies is the $S_2(0):(2,2)\rightarrow(0,0)$ transition in H$_2$ \cite{Cozijn:23}. The theoretical prediction of $252\,016\,361.45(69)$ MHz~\cite{PK:25a}, which already incorporates this correction, closely aligns with the experimental value of $252\,016\,361.164(8)$ MHz. The observed difference of $0.28(69)$ MHz is comparable to the contribution of this correction alone, which is $0.19$ MHz. 

These two examples clearly illustrate the importance of the electron-nucleus Araki-Sucher correction in achieving the final accuracy. Similar observations can be made for the other energy levels and transitions. 

\begin{table*}[ht]
\caption{Contribution of the $E^{(5)}_\mathrm{AS,en}$ (in $10^{-6}\,\icm$) to the dissociation energy
of rovibrational levels with the lowest quantum numbers $(v,J)$. Uncertainties are estimated by scaling each entry by the factor $1/\mun$.}
\label{Tab:ASendiss}
\begin{tabular*}{\textwidth}{c@{\extracolsep{\fill}}c*{6}{x{3.7}}}
\hline\hline
$v$&$J$& \centt{H$_2$}& \centt{HD}   & \centt{HT}   & \centt{D$_2$}& \centt{DT}   & \centt{T$_2$}\\
\hline
0 & 0 & -28.677(31) & -21.688(18) & -19.344(14) & -14.604(8) & -12.223(6) & -9.828(4) \\
0 & 1 & -28.489(31) & -21.582(18) & -19.260(14) & -14.557(8) & -12.190(6) & -9.807(4) \\
0 & 2 & -28.117(31) & -21.371(17) & -19.093(14) & -14.462(8) & -12.124(6) & -9.765(4) \\
0 & 3 & -27.568(30) & -21.059(17) & -18.844(14) & -14.322(8) & -12.026(5) & -9.702(4) \\
0 & 4 & -26.850(29) & -20.649(17) & -18.518(13) & -14.136(8) & -11.896(5) & -9.618(4) \\
0 & 5 & -25.976(28) & -20.146(16) & -18.118(13) & -13.907(8) & -11.736(5) & -9.514(3) \\
0 & 6 & -24.960(27) & -19.557(16) & -17.647(13) & -13.637(7) & -11.546(5) & -9.391(3) \\
1 & 0 & -25.332(28) & -19.494(16) & -17.498(13) & -13.396(7) & -11.299(5) & -9.163(3) \\
1 & 1 & -25.155(27) & -19.394(16) & -17.417(13) & -13.350(7) & -11.267(5) & -9.142(3) \\
1 & 2 & -24.805(27) & -19.193(16) & -17.258(13) & -13.260(7) & -11.204(5) & -9.101(3) \\
1 & 3 & -24.287(26) & -18.896(15) & -17.022(12) & -13.125(7) & -11.109(5) & -9.040(3) \\
1 & 4 & -23.611(26) & -18.507(15) & -16.711(12) & -12.947(7) & -10.984(5) & -8.959(3) \\
1 & 5 & -22.788(25) & -18.029(15) & -16.329(12) & -12.727(7) & -10.830(5) & -8.859(3) \\
1 & 6 & -21.832(24) & -17.470(14) & -15.881(12) & -12.468(7) & -10.647(5) & -8.740(3) \\
2 & 0 & -22.221(24) & -17.433(14) & -15.757(11) & -12.247(7) & -10.417(5) & -8.524(3) \\
2 & 1 & -22.054(24) & -17.338(14) & -15.681(11) & -12.204(7) & -10.386(5) & -8.504(3) \\
2 & 2 & -21.724(24) & -17.147(14) & -15.529(11) & -12.117(7) & -10.325(5) & -8.465(3) \\
2 & 3 & -21.237(23) & -16.865(14) & -15.304(11) & -11.987(7) & -10.234(5) & -8.406(3) \\
2 & 4 & -20.601(22) & -16.496(13) & -15.008(11) & -11.817(6) & -10.114(5) & -8.328(3) \\
2 & 5 & -19.827(22) & -16.043(13) & -14.645(11) & -11.606(6) & -9.965(5)  & -8.231(3) \\
2 & 6 & -18.929(21) & -15.513(13) & -14.219(10) & -11.358(6) & -9.790(4)  & -8.116(3) \\
3 & 0 & -19.329(21) & -15.498(13) & -14.116(10) & -11.156(6) & -9.575(4)  & -7.912(3) \\
3 & 1 & -19.172(21) & -15.407(13) & -14.044(10) & -11.114(6) & -9.545(4)  & -7.893(3) \\
3 & 2 & -18.862(21) & -15.227(12) & -13.900(10) & -11.031(6) & -9.486(4)  & -7.855(3) \\
3 & 3 & -18.404(20) & -14.959(12) & -13.685(10) & -10.907(6) & -9.399(4)  & -7.797(3) \\
3 & 4 & -17.806(19) & -14.609(12) & -13.404(10) & -10.743(6) & -9.283(4)  & -7.722(3) \\
3 & 5 & -17.081(19) & -14.180(12) & -13.059(9)  & -10.542(6) & -9.140(4)  & -7.628(3) \\
3 & 6 & -16.238(18) & -13.678(11) & -12.654(9)  & -10.304(6) & -8.971(4)  & -7.517(3) \\
\hline\hline
\end{tabular*} 
\end{table*}

\section{Conclusion}

The results presented in this work demonstrate that the electron–nucleus Araki–Sucher correction, although suppressed by the electron-to-nucleus mass ratio, contributes non-negligibly to the total QED correction in molecular hydrogen and its isotopologues. 

The magnitude of the correction do dissociation energy, on the order of $10^{-5}\,\icm$, is comparable to the current total theoretical uncertainty. This highlights the importance of including the Araki–Sucher term in theoretical models aiming to match modern experimental precision.

The correction is largest for H$_2$ and decreases systematically with increasing nuclear mass, reaching its smallest value for T$_2$. This behavior confirms the theoretical scaling of the electron–nucleus Araki–Sucher term and validates the computational approach employed.

It is worth noting that the present calculations were performed within the Born–Oppenheimer approximation, neglecting nonadiabatic effects. While the numerical uncertainty due to basis set and integration convergence is well controlled, the dominant source of error arises from the BO approximation itself. An estimate of this uncertainty, based on the scaling $\me/\mun\,{E}^{(5)}_{\mathrm{AS,en}}$, suggests a relative error of approximately 0.1\%, which is acceptable for the current level of theoretical-experimental comparison.

Future work may include the extension of this approach to incorporate nonadiabatic corrections and to evaluate the Araki–Sucher term beyond the clamped-nuclei framework. Such developments would be essential for pushing further the theoretical accuracy.

\section*{Acknowledgment}

The author thanks Krzysztof Pachucki for the valuable discussions that contributed to this work. This research was supported by the National Science Center (Poland) Grant No. 2021/41/B/ST4/00089. A~computer grant from the Poznań Supercomputing and Networking Center was used to carry out the numerical calculations. 

\bibliography{AS}

\begin{thebibliography}{24}%
\makeatletter
\providecommand \@ifxundefined [1]{%
 \@ifx{#1\undefined}
}%
\providecommand \@ifnum [1]{%
 \ifnum #1\expandafter \@firstoftwo
 \else \expandafter \@secondoftwo
 \fi
}%
\providecommand \@ifx [1]{%
 \ifx #1\expandafter \@firstoftwo
 \else \expandafter \@secondoftwo
 \fi
}%
\providecommand \natexlab [1]{#1}%
\providecommand \enquote  [1]{``#1''}%
\providecommand \bibnamefont  [1]{#1}%
\providecommand \bibfnamefont [1]{#1}%
\providecommand \citenamefont [1]{#1}%
\providecommand \href@noop [0]{\@secondoftwo}%
\providecommand \href [0]{\begingroup \@sanitize@url \@href}%
\providecommand \@href[1]{\@@startlink{#1}\@@href}%
\providecommand \@@href[1]{\endgroup#1\@@endlink}%
\providecommand \@sanitize@url [0]{\catcode `\\12\catcode `\$12\catcode
  `\&12\catcode `\#12\catcode `\^12\catcode `\_12\catcode `\%12\relax}%
\providecommand \@@startlink[1]{}%
\providecommand \@@endlink[0]{}%
\providecommand \url  [0]{\begingroup\@sanitize@url \@url }%
\providecommand \@url [1]{\endgroup\@href {#1}{\urlprefix }}%
\providecommand \urlprefix  [0]{URL }%
\providecommand \Eprint [0]{\href }%
\providecommand \doibase [0]{https://doi.org/}%
\providecommand \selectlanguage [0]{\@gobble}%
\providecommand \bibinfo  [0]{\@secondoftwo}%
\providecommand \bibfield  [0]{\@secondoftwo}%
\providecommand \translation [1]{[#1]}%
\providecommand \BibitemOpen [0]{}%
\providecommand \bibitemStop [0]{}%
\providecommand \bibitemNoStop [0]{.\EOS\space}%
\providecommand \EOS [0]{\spacefactor3000\relax}%
\providecommand \BibitemShut  [1]{\csname bibitem#1\endcsname}%
\let\auto@bib@innerbib\@empty
\bibitem [{\citenamefont {Fast}\ and\ \citenamefont {Meek}(2020)}]{Fast:20}%
  \BibitemOpen
  \bibfield  {author} {\bibinfo {author} {\bibfnamefont {A.}~\bibnamefont
  {Fast}}\ and\ \bibinfo {author} {\bibfnamefont {S.~A.}\ \bibnamefont
  {Meek}},\ }\bibfield  {title} {\bibinfo {title} {{Sub-ppb Measurement of a
  Fundamental Band Rovibrational Transition in {HD}}},\ }\href
  {https://doi.org/10.1103/PhysRevLett.125.023001} {\bibfield  {journal}
  {\bibinfo  {journal} {Phys. Rev. Lett.}\ }\textbf {\bibinfo {volume} {125}},\
  \bibinfo {pages} {023001} (\bibinfo {year} {2020})}\BibitemShut {NoStop}%
\bibitem [{\citenamefont {Diouf}\ \emph {et~al.}(2020)\citenamefont {Diouf},
  \citenamefont {Cozijn}, \citenamefont {Lai}, \citenamefont {Salumbides},\
  and\ \citenamefont {Ubachs}}]{Diouf:20}%
  \BibitemOpen
  \bibfield  {author} {\bibinfo {author} {\bibfnamefont {M.~L.}\ \bibnamefont
  {Diouf}}, \bibinfo {author} {\bibfnamefont {F.~M.~J.}\ \bibnamefont
  {Cozijn}}, \bibinfo {author} {\bibfnamefont {K.-F.}\ \bibnamefont {Lai}},
  \bibinfo {author} {\bibfnamefont {E.~J.}\ \bibnamefont {Salumbides}},\ and\
  \bibinfo {author} {\bibfnamefont {W.}~\bibnamefont {Ubachs}},\ }\bibfield
  {title} {\bibinfo {title} {Lamb-peak spectrum of the {HD} (2-0) {$P(1)$}
  line},\ }\href {https://doi.org/10.1103/PhysRevResearch.2.023209} {\bibfield
  {journal} {\bibinfo  {journal} {Phys. Rev. Research}\ }\textbf {\bibinfo
  {volume} {2}},\ \bibinfo {pages} {023209} (\bibinfo {year}
  {2020})}\BibitemShut {NoStop}%
\bibitem [{\citenamefont {Fast}\ and\ \citenamefont {Meek}(2022)}]{Fast:22}%
  \BibitemOpen
  \bibfield  {author} {\bibinfo {author} {\bibfnamefont {A.}~\bibnamefont
  {Fast}}\ and\ \bibinfo {author} {\bibfnamefont {S.~A.}\ \bibnamefont
  {Meek}},\ }\bibfield  {title} {\bibinfo {title} {Precise measurement of the
  {D$_2$} {S$_1$(0)} vibrational transition frequency},\ }\href
  {https://doi.org/10.1080/00268976.2021.1999520} {\bibfield  {journal}
  {\bibinfo  {journal} {Mol. Phys.}\ }\textbf {\bibinfo {volume} {120}},\
  \bibinfo {pages} {e1999520} (\bibinfo {year} {2022})}\BibitemShut {NoStop}%
\bibitem [{\citenamefont {Cozijn}\ \emph {et~al.}(2023)\citenamefont {Cozijn},
  \citenamefont {Diouf},\ and\ \citenamefont {Ubachs}}]{Cozijn:23}%
  \BibitemOpen
  \bibfield  {author} {\bibinfo {author} {\bibfnamefont {F.~M.~J.}\
  \bibnamefont {Cozijn}}, \bibinfo {author} {\bibfnamefont {M.~L.}\
  \bibnamefont {Diouf}},\ and\ \bibinfo {author} {\bibfnamefont
  {W.}~\bibnamefont {Ubachs}},\ }\bibfield  {title} {\bibinfo {title} {{Lamb
  Dip of a Quadrupole Transition in ${\mathrm{H}}_{2}$}},\ }\href
  {https://doi.org/10.1103/PhysRevLett.131.073001} {\bibfield  {journal}
  {\bibinfo  {journal} {Phys. Rev. Lett.}\ }\textbf {\bibinfo {volume} {131}},\
  \bibinfo {pages} {073001} (\bibinfo {year} {2023})}\BibitemShut {NoStop}%
\bibitem [{\citenamefont {Diouf}\ \emph {et~al.}(2024)\citenamefont {Diouf},
  \citenamefont {Cozijn},\ and\ \citenamefont {Ubachs}}]{Diouf:24}%
  \BibitemOpen
  \bibfield  {author} {\bibinfo {author} {\bibfnamefont {M.~L.}\ \bibnamefont
  {Diouf}}, \bibinfo {author} {\bibfnamefont {F.~M.~J.}\ \bibnamefont
  {Cozijn}},\ and\ \bibinfo {author} {\bibfnamefont {W.}~\bibnamefont
  {Ubachs}},\ }\bibfield  {title} {\bibinfo {title} {Hyperfine structure in a
  vibrational quadrupole transition of ortho-{H$_2$}},\ }\href
  {https://doi.org/10.1080/00268976.2024.2304101} {\bibfield  {journal}
  {\bibinfo  {journal} {Mol. Phys.}\ }\textbf {\bibinfo {volume} {122}},\
  \bibinfo {pages} {e2304101} (\bibinfo {year} {2024})}\BibitemShut {NoStop}%
\bibitem [{\citenamefont {Cozijn}\ \emph {et~al.}(2024)\citenamefont {Cozijn},
  \citenamefont {Diouf}, \citenamefont {Ubachs}, \citenamefont {Hermann},\ and\
  \citenamefont {Schl\"osser}}]{Cozijn:24}%
  \BibitemOpen
  \bibfield  {author} {\bibinfo {author} {\bibfnamefont {F.~M.~J.}\
  \bibnamefont {Cozijn}}, \bibinfo {author} {\bibfnamefont {M.~L.}\
  \bibnamefont {Diouf}}, \bibinfo {author} {\bibfnamefont {W.}~\bibnamefont
  {Ubachs}}, \bibinfo {author} {\bibfnamefont {V.}~\bibnamefont {Hermann}},\
  and\ \bibinfo {author} {\bibfnamefont {M.}~\bibnamefont {Schl\"osser}},\
  }\bibfield  {title} {\bibinfo {title} {Precision measurement of vibrational
  quanta in tritium hydride},\ }\href
  {https://doi.org/10.1103/PhysRevLett.132.113002} {\bibfield  {journal}
  {\bibinfo  {journal} {Phys. Rev. Lett.}\ }\textbf {\bibinfo {volume} {132}},\
  \bibinfo {pages} {113002} (\bibinfo {year} {2024})}\BibitemShut {NoStop}%
\bibitem [{\citenamefont {Stankiewicz}\ \emph {et~al.}(2025)\citenamefont
  {Stankiewicz}, \citenamefont {Makowski}, \citenamefont {Słowiński},
  \citenamefont {Sołtys}, \citenamefont {Bednarski}, \citenamefont
  {Jóźwiak}, \citenamefont {Stolarczyk}, \citenamefont {Narożnik},
  \citenamefont {Kierski}, \citenamefont {Wójtewicz}, \citenamefont {Cygan},
  \citenamefont {Kowzan}, \citenamefont {Masłowski}, \citenamefont
  {Piwiński}, \citenamefont {Lisak},\ and\ \citenamefont
  {Wcisło}}]{Stankiewicz:25}%
  \BibitemOpen
  \bibfield  {author} {\bibinfo {author} {\bibfnamefont {K.}~\bibnamefont
  {Stankiewicz}}, \bibinfo {author} {\bibfnamefont {M.}~\bibnamefont
  {Makowski}}, \bibinfo {author} {\bibfnamefont {M.}~\bibnamefont
  {Słowiński}}, \bibinfo {author} {\bibfnamefont {K.~L.}\ \bibnamefont
  {Sołtys}}, \bibinfo {author} {\bibfnamefont {B.}~\bibnamefont {Bednarski}},
  \bibinfo {author} {\bibfnamefont {H.}~\bibnamefont {Jóźwiak}}, \bibinfo
  {author} {\bibfnamefont {N.}~\bibnamefont {Stolarczyk}}, \bibinfo {author}
  {\bibfnamefont {M.}~\bibnamefont {Narożnik}}, \bibinfo {author}
  {\bibfnamefont {D.}~\bibnamefont {Kierski}}, \bibinfo {author} {\bibfnamefont
  {S.}~\bibnamefont {Wójtewicz}}, \bibinfo {author} {\bibfnamefont
  {A.}~\bibnamefont {Cygan}}, \bibinfo {author} {\bibfnamefont
  {G.}~\bibnamefont {Kowzan}}, \bibinfo {author} {\bibfnamefont
  {P.}~\bibnamefont {Masłowski}}, \bibinfo {author} {\bibfnamefont
  {M.}~\bibnamefont {Piwiński}}, \bibinfo {author} {\bibfnamefont
  {D.}~\bibnamefont {Lisak}},\ and\ \bibinfo {author} {\bibfnamefont
  {P.}~\bibnamefont {Wcisło}},\ }\href {https://arxiv.org/abs/2502.12703}
  {\bibinfo {title} {Cavity-enhanced spectroscopy in the deep cryogenic regime
  -- new hydrogen technologies for quantum sensing}} (\bibinfo {year} {2025}),\
  \Eprint {https://arxiv.org/abs/2502.12703} {arXiv:2502.12703
  [physics.atom-ph]} \BibitemShut {NoStop}%
\bibitem [{\citenamefont {Araki}(1957)}]{Araki:57}%
  \BibitemOpen
  \bibfield  {author} {\bibinfo {author} {\bibfnamefont {H.}~\bibnamefont
  {Araki}},\ }\bibfield  {title} {\bibinfo {title} {{Quantum-Electrodynamical
  Corrections to Energy-Levels of Helium}},\ }\href
  {https://doi.org/10.1143/PTP.17.619} {\bibfield  {journal} {\bibinfo
  {journal} {Prog. Theor. Phys.}\ }\textbf {\bibinfo {volume} {17}},\ \bibinfo
  {pages} {619} (\bibinfo {year} {1957})}\BibitemShut {NoStop}%
\bibitem [{\citenamefont {Sucher}(1958)}]{Sucher:58}%
  \BibitemOpen
  \bibfield  {author} {\bibinfo {author} {\bibfnamefont {J.}~\bibnamefont
  {Sucher}},\ }\bibfield  {title} {\bibinfo {title} {{Energy Levels of the
  Two-Electron Atom to Order ${\ensuremath{\alpha}}^{3}$ ry; Ionization Energy
  of Helium}},\ }\href {https://doi.org/10.1103/PhysRev.109.1010} {\bibfield
  {journal} {\bibinfo  {journal} {Phys. Rev.}\ }\textbf {\bibinfo {volume}
  {109}},\ \bibinfo {pages} {1010} (\bibinfo {year} {1958})}\BibitemShut
  {NoStop}%
\bibitem [{\citenamefont {Pachucki}\ \emph {et~al.}(2005)\citenamefont
  {Pachucki}, \citenamefont {Cencek},\ and\ \citenamefont {Komasa}}]{PCK:05}%
  \BibitemOpen
  \bibfield  {author} {\bibinfo {author} {\bibfnamefont {K.}~\bibnamefont
  {Pachucki}}, \bibinfo {author} {\bibfnamefont {W.}~\bibnamefont {Cencek}},\
  and\ \bibinfo {author} {\bibfnamefont {J.}~\bibnamefont {Komasa}},\
  }\bibfield  {title} {\bibinfo {title} {{On the acceleration of the
  convergence of singular operators in Gaussian basis sets}},\ }\href
  {https://doi.org/10.1063/1.1888572} {\bibfield  {journal} {\bibinfo
  {journal} {J. Chem. Phys.}\ }\textbf {\bibinfo {volume} {122}},\ \bibinfo
  {eid} {184101} (\bibinfo {year} {2005})}\BibitemShut {NoStop}%
\bibitem [{\citenamefont {Korobov}(2006)}]{Korobov:06}%
  \BibitemOpen
  \bibfield  {author} {\bibinfo {author} {\bibfnamefont {V.~I.}\ \bibnamefont
  {Korobov}},\ }\bibfield  {title} {\bibinfo {title} {{Leading-order
  relativistic and radiative corrections to the rovibrational spectrum of
  $\mathrm{H}_{2}^{+}$ and $\mathrm{H}{\mathrm{D}}^{+}$ molecular ions}},\
  }\href {https://doi.org/10.1103/PhysRevA.74.052506} {\bibfield  {journal}
  {\bibinfo  {journal} {Phys. Rev. A}\ }\textbf {\bibinfo {volume} {74}},\
  \bibinfo {pages} {052506} (\bibinfo {year} {2006})}\BibitemShut {NoStop}%
\bibitem [{\citenamefont {Piszczatowski}\ \emph {et~al.}(2009)\citenamefont
  {Piszczatowski}, \citenamefont {Lach}, \citenamefont {Przybytek},
  \citenamefont {Komasa}, \citenamefont {Pachucki},\ and\ \citenamefont
  {Jeziorski}}]{Piszczatowski:09}%
  \BibitemOpen
  \bibfield  {author} {\bibinfo {author} {\bibfnamefont {K.}~\bibnamefont
  {Piszczatowski}}, \bibinfo {author} {\bibfnamefont {G.}~\bibnamefont {Lach}},
  \bibinfo {author} {\bibfnamefont {M.}~\bibnamefont {Przybytek}}, \bibinfo
  {author} {\bibfnamefont {J.}~\bibnamefont {Komasa}}, \bibinfo {author}
  {\bibfnamefont {K.}~\bibnamefont {Pachucki}},\ and\ \bibinfo {author}
  {\bibfnamefont {B.}~\bibnamefont {Jeziorski}},\ }\bibfield  {title} {\bibinfo
  {title} {Theoretical determination of the dissociation energy of molecular
  hydrogen},\ }\href {https://doi.org/10.1021/ct900391p} {\bibfield  {journal}
  {\bibinfo  {journal} {J. Chem. Theory Comput.}\ }\textbf {\bibinfo {volume}
  {5}},\ \bibinfo {pages} {3039} (\bibinfo {year} {2009})}\BibitemShut
  {NoStop}%
\bibitem [{\citenamefont {Pachucki}\ and\ \citenamefont
  {Komasa}(2010)}]{PK:10}%
  \BibitemOpen
  \bibfield  {author} {\bibinfo {author} {\bibfnamefont {K.}~\bibnamefont
  {Pachucki}}\ and\ \bibinfo {author} {\bibfnamefont {J.}~\bibnamefont
  {Komasa}},\ }\bibfield  {title} {\bibinfo {title} {{Rovibrational levels of
  HD}},\ }\href {https://doi.org/10.1039/C0CP00209G} {\bibfield  {journal}
  {\bibinfo  {journal} {Phys. Chem. Chem. Phys.}\ }\textbf {\bibinfo {volume}
  {12}},\ \bibinfo {pages} {9188} (\bibinfo {year} {2010})}\BibitemShut
  {NoStop}%
\bibitem [{\citenamefont {Komasa}\ \emph {et~al.}(2011)\citenamefont {Komasa},
  \citenamefont {Piszczatowski}, \citenamefont {Lach}, \citenamefont
  {Przybytek}, \citenamefont {Jeziorski},\ and\ \citenamefont
  {Pachucki}}]{Komasa:11}%
  \BibitemOpen
  \bibfield  {author} {\bibinfo {author} {\bibfnamefont {J.}~\bibnamefont
  {Komasa}}, \bibinfo {author} {\bibfnamefont {K.}~\bibnamefont
  {Piszczatowski}}, \bibinfo {author} {\bibfnamefont {G.}~\bibnamefont {Lach}},
  \bibinfo {author} {\bibfnamefont {M.}~\bibnamefont {Przybytek}}, \bibinfo
  {author} {\bibfnamefont {B.}~\bibnamefont {Jeziorski}},\ and\ \bibinfo
  {author} {\bibfnamefont {K.}~\bibnamefont {Pachucki}},\ }\bibfield  {title}
  {\bibinfo {title} {{Quantum Electrodynamics Effects in Rovibrational Spectra
  of Molecular Hydrogen}},\ }\href {https://doi.org/10.1021/ct200438t}
  {\bibfield  {journal} {\bibinfo  {journal} {J. Chem. Theory Comput.}\
  }\textbf {\bibinfo {volume} {7}},\ \bibinfo {pages} {3105} (\bibinfo {year}
  {2011})}\BibitemShut {NoStop}%
\bibitem [{\citenamefont {Balcerzak}\ \emph {et~al.}(2017)\citenamefont
  {Balcerzak}, \citenamefont {Lesiuk},\ and\ \citenamefont
  {Moszynski}}]{Balcerzak:17}%
  \BibitemOpen
  \bibfield  {author} {\bibinfo {author} {\bibfnamefont {J.~G.}\ \bibnamefont
  {Balcerzak}}, \bibinfo {author} {\bibfnamefont {M.}~\bibnamefont {Lesiuk}},\
  and\ \bibinfo {author} {\bibfnamefont {R.}~\bibnamefont {Moszynski}},\
  }\bibfield  {title} {\bibinfo {title} {{Calculation of Araki-Sucher
  correction for many-electron systems}},\ }\href
  {https://doi.org/10.1103/PhysRevA.96.052510} {\bibfield  {journal} {\bibinfo
  {journal} {Phys. Rev. A}\ }\textbf {\bibinfo {volume} {96}},\ \bibinfo
  {pages} {052510} (\bibinfo {year} {2017})}\BibitemShut {NoStop}%
\bibitem [{\citenamefont {Stanke}\ \emph {et~al.}(2017)\citenamefont {Stanke},
  \citenamefont {Jurkowski},\ and\ \citenamefont {Adamowicz}}]{Stanke:17}%
  \BibitemOpen
  \bibfield  {author} {\bibinfo {author} {\bibfnamefont {M.}~\bibnamefont
  {Stanke}}, \bibinfo {author} {\bibfnamefont {J.}~\bibnamefont {Jurkowski}},\
  and\ \bibinfo {author} {\bibfnamefont {L.}~\bibnamefont {Adamowicz}},\
  }\bibfield  {title} {\bibinfo {title} {{Algorithms for calculating the
  leading quantum electrodynamics $P(1/r^3)$ correction with all-electron
  molecular explicitly correlated Gaussians}},\ }\href
  {https://doi.org/10.1088/1361-6455/aa56ad} {\bibfield  {journal} {\bibinfo
  {journal} {J. Phys. B}\ }\textbf {\bibinfo {volume} {50}},\ \bibinfo {pages}
  {065101} (\bibinfo {year} {2017})}\BibitemShut {NoStop}%
\bibitem [{\citenamefont {Si\l{}kowski}\ \emph {et~al.}(2023)\citenamefont
  {Si\l{}kowski}, \citenamefont {Pachucki}, \citenamefont {Komasa},\ and\
  \citenamefont {Puchalski}}]{SPKP:23}%
  \BibitemOpen
  \bibfield  {author} {\bibinfo {author} {\bibfnamefont {M.}~\bibnamefont
  {Si\l{}kowski}}, \bibinfo {author} {\bibfnamefont {K.}~\bibnamefont
  {Pachucki}}, \bibinfo {author} {\bibfnamefont {J.}~\bibnamefont {Komasa}},\
  and\ \bibinfo {author} {\bibfnamefont {M.}~\bibnamefont {Puchalski}},\
  }\bibfield  {title} {\bibinfo {title} {Leading-order {QED} effects in the
  ground electronic state of molecular hydrogen},\ }\href
  {https://doi.org/10.1103/PhysRevA.107.032807} {\bibfield  {journal} {\bibinfo
   {journal} {Phys. Rev. A}\ }\textbf {\bibinfo {volume} {107}},\ \bibinfo
  {pages} {032807} (\bibinfo {year} {2023})}\BibitemShut {NoStop}%
\bibitem [{\citenamefont {Puchalski}\ \emph
  {et~al.}(2019{\natexlab{a}})\citenamefont {Puchalski}, \citenamefont
  {Komasa}, \citenamefont {Czachorowski},\ and\ \citenamefont
  {Pachucki}}]{PKCP:19}%
  \BibitemOpen
  \bibfield  {author} {\bibinfo {author} {\bibfnamefont {M.}~\bibnamefont
  {Puchalski}}, \bibinfo {author} {\bibfnamefont {J.}~\bibnamefont {Komasa}},
  \bibinfo {author} {\bibfnamefont {P.}~\bibnamefont {Czachorowski}},\ and\
  \bibinfo {author} {\bibfnamefont {K.}~\bibnamefont {Pachucki}},\ }\bibfield
  {title} {\bibinfo {title} {{Nonadiabatic QED Correction to the Dissociation
  Energy of the Hydrogen Molecule}},\ }\href
  {https://doi.org/10.1103/PhysRevLett.122.103003} {\bibfield  {journal}
  {\bibinfo  {journal} {Phys. Rev. Lett.}\ }\textbf {\bibinfo {volume} {122}},\
  \bibinfo {pages} {103003} (\bibinfo {year} {2019}{\natexlab{a}})}\BibitemShut
  {NoStop}%
\bibitem [{\citenamefont {Puchalski}\ \emph
  {et~al.}(2019{\natexlab{b}})\citenamefont {Puchalski}, \citenamefont
  {Komasa}, \citenamefont {Spyszkiewicz},\ and\ \citenamefont
  {Pachucki}}]{PKSP:19}%
  \BibitemOpen
  \bibfield  {author} {\bibinfo {author} {\bibfnamefont {M.}~\bibnamefont
  {Puchalski}}, \bibinfo {author} {\bibfnamefont {J.}~\bibnamefont {Komasa}},
  \bibinfo {author} {\bibfnamefont {A.}~\bibnamefont {Spyszkiewicz}},\ and\
  \bibinfo {author} {\bibfnamefont {K.}~\bibnamefont {Pachucki}},\ }\bibfield
  {title} {\bibinfo {title} {Dissociation energy of molecular hydrogen
  isotopologues},\ }\href {https://doi.org/10.1103/PhysRevA.100.020503}
  {\bibfield  {journal} {\bibinfo  {journal} {Phys. Rev. A}\ }\textbf {\bibinfo
  {volume} {100}},\ \bibinfo {pages} {020503} (\bibinfo {year}
  {2019}{\natexlab{b}})}\BibitemShut {NoStop}%
\bibitem [{\citenamefont {Pachucki}\ and\ \citenamefont
  {Komasa}(2025)}]{PK:25a}%
  \BibitemOpen
  \bibfield  {author} {\bibinfo {author} {\bibfnamefont {K.}~\bibnamefont
  {Pachucki}}\ and\ \bibinfo {author} {\bibfnamefont {J.}~\bibnamefont
  {Komasa}},\ }\bibfield  {title} {\bibinfo {title} {{From First Principles to
  Quantum Electrodynamics: Pushing the Limits of Theory with the Hydrogen
  Molecule}}} (\bibinfo {year} {2025}),\ \bibinfo {note} {accepted for
  publication in J. Chem. Theory Comput.}\BibitemShut {Stop}%
\bibitem [{\citenamefont {Singer}(1960)}]{Singer:60}%
  \BibitemOpen
  \bibfield  {author} {\bibinfo {author} {\bibfnamefont {K.}~\bibnamefont
  {Singer}},\ }\bibfield  {title} {\bibinfo {title} {{The Use of Gaussian
  (Exponential Quadratic) Wave Functions in Molecular Problems. I. General
  Formulae for the Evaluation of Integrals}},\ }\href
  {https://doi.org/10.1098/rspa.1960.0196} {\bibfield  {journal} {\bibinfo
  {journal} {Proc. R. Soc. Lond. A}\ }\textbf {\bibinfo {volume} {258}},\
  \bibinfo {pages} {412} (\bibinfo {year} {1960})}\BibitemShut {NoStop}%
\bibitem [{\citenamefont {Pachucki}\ \emph {et~al.}(2014)\citenamefont
  {Pachucki}, \citenamefont {Puchalski},\ and\ \citenamefont
  {Yerokhin}}]{GAUSEXT}%
  \BibitemOpen
  \bibfield  {author} {\bibinfo {author} {\bibfnamefont {K.}~\bibnamefont
  {Pachucki}}, \bibinfo {author} {\bibfnamefont {M.}~\bibnamefont
  {Puchalski}},\ and\ \bibinfo {author} {\bibfnamefont {V.}~\bibnamefont
  {Yerokhin}},\ }\bibfield  {title} {\bibinfo {title} {{Extended Gaussian
  quadratures for functions with an end-point singularity of logarithmic
  type}},\ }\href {https://doi.org/https://doi.org/10.1016/j.cpc.2014.06.018}
  {\bibfield  {journal} {\bibinfo  {journal} {Comput. Phys. Commun.}\ }\textbf
  {\bibinfo {volume} {185}},\ \bibinfo {pages} {2913} (\bibinfo {year}
  {2014})}\BibitemShut {NoStop}%
\bibitem [{\citenamefont {{H2Spectre ver. 7.4 F}ortran~source
  code}(2022)}]{H2Spectre}%
  \BibitemOpen
  \bibfield  {author} {\bibinfo {author} {\bibnamefont {{H2Spectre ver. 7.4
  F}ortran~source code}},\ }\href@noop {} {} (\bibinfo {year} {2022}),\
  \bibinfo {note}
  {{\url{https://qcg.home.amu.edu.pl/H2Spectre.html}}}\BibitemShut {NoStop}%
\bibitem [{\citenamefont {Mohr}\ \emph {et~al.}(2025)\citenamefont {Mohr},
  \citenamefont {Newell}, \citenamefont {Taylor},\ and\ \citenamefont
  {Tiesinga}}]{CODATA:22}%
  \BibitemOpen
  \bibfield  {author} {\bibinfo {author} {\bibfnamefont {P.~J.}\ \bibnamefont
  {Mohr}}, \bibinfo {author} {\bibfnamefont {D.~B.}\ \bibnamefont {Newell}},
  \bibinfo {author} {\bibfnamefont {B.~N.}\ \bibnamefont {Taylor}},\ and\
  \bibinfo {author} {\bibfnamefont {E.}~\bibnamefont {Tiesinga}},\ }\bibfield
  {title} {\bibinfo {title} {Codata recommended values of the fundamental
  physical constants: 2022},\ }\href
  {https://doi.org/10.1103/RevModPhys.97.025002} {\bibfield  {journal}
  {\bibinfo  {journal} {Rev. Mod. Phys.}\ }\textbf {\bibinfo {volume} {97}},\
  \bibinfo {pages} {025002} (\bibinfo {year} {2025})}\BibitemShut {NoStop}%
\end{thebibliography}%

\end{document}